\documentclass[11pt,twoside]{article}  
\usepackage{apn3conf}

\begin{document}   

%
%
%
%

\title{Magellanic Cloud planetary
nebulae as probes of stellar evolution and populations}

%
%
%

\author{Letizia Stanghellini}
\affil{Space Telescope Science Institute, and European Space Agency}

%
%

\contact{Letizia Stanghellini}
\email{lstanghe@stsci.edu}

%
%
%
%
%

\paindex{Stanghellini, L.}

%
%

\authormark{STANGHELLINI}

%
%



\begin{abstract}          
Magellanic Cloud Planetary Nebulae (PNs) offer insight 
of both the population and evolution of low- and 
intermediate-mass stars, in environments that are free of the 
distance bias and the differential reddening that hinder the
observations of the Galactic sample. The study of LMC and SMC PNs also
offers the direct comparison of stellar populations with different
metallicity. 
We present a selection of the results from our recent {\it HST}
surveys, including (1) 
the morphological analysis of Magellanic PNs, and the statistics
of the morphological samples in the LMC and the SMC; (2) 
the surface brightness versus radius relationship; and (3)
the
analysis and modeling of the [O III]/H$\beta$ PN luminosity functions in the
LMC and the SMC.

\end{abstract}

%
%

\section{Introduction}

Planetary Nebulae (PNs) are important probes of stellar evolution, stellar
populations, and cosmic recycling. PNs have been observed
in the Local Group as well as in external galaxies, probing stellar evolution and 
populations in relation to their environment.

The details of the observations of Galactic PNs and their central stars (CSs)
typically surpass the 
details of stellar and hydrodynamic models. Galactic PN studies are a
necessary background
toward the understanding the PN populations in general. 
Yet, the distance scale of Galactic PNs is uncertain to such a degree that 
the meaning of the comparison between observations and theory is hindered. 
By the same token, statistical studies 
of PN populations in the Galaxy suffer for the observational
bias against the detection of Galactic disk PNs, and for the patchy interstellar
extinction. 

PNs in the Magellanic Clouds (LMC, SMC), hundreds of low-extinction 
planetaries at uniformly known distances, are a real bounty for the stellar
evolution scientist. The composition gradient between the LMC, the SMC,
and the Galaxy, afford the study of the effects
of environment metallicity on PN evolution. The relative vicinity of the Clouds,
and the spatial resolution that can be achieved with the {\it Hubble Space Telescope (HST)}, 
allow the detection of PN morphology. 
Studying the PNs in the
Magellanic Clouds is a necessary step toward the understanding of the 
onset of morphological type and its relation to metallicity and stellar evolution.

\section{Our Magellanic PN program}

During the {\it HST} Cycle 8 we started a series of surveys aimed at obtain the 
size, morphology, and CS properties of all Magellanic PNs known to date.
The {\it HST} was an obligatory choice, since the Magellanic PNs are typically
half an arcsecond across, thus they are generally not resolved with ground-based
telescopes. 

The medium-dispersion, slitless capability of STIS offers us a valuable 
opportunity to study the evolution and morphology of the Magellanic
Cloud PNs and 
their CSs at once.  We have applied this capability in several
SNAPSHOT surveys, obtaining images in the light of up to 7 of 
the most prominent low- and moderate-ionization optical, nebular emission 
lines ({\it HST} programs 8271, 8663, and 9077).  We also obtained direct continuum 
images to identify the correct CS (in spite of the crowded fields), and 
to measure the optical continuum emission.

In addition to the optical slitless spectra and broad band continuum images
of the LMC and SMC PNs, we have acquired STIS UV spectra of 24 LMC PNs.
In the cases where the CSs were hard to find in our
STIS broad band images, we have also
acquired WFPC2 Str\"omgren images (Program 8702).
We have used through our investigations the limited data available
in the literature (see Stanghellini et al. 1999). 
The data acquired by us with {\it HST} 
can be easily retrieved from the dedicated MAST page
\htmladdURL{http://archive.stsci.edu/hst/mcpn}. 
In addition to {\it HST}, we
have made extensive use of the spectra acquired from the ground by
us (papers in preparation), and available in the literature.

\begin{figure}
\epsscale{.40}
\plottwo{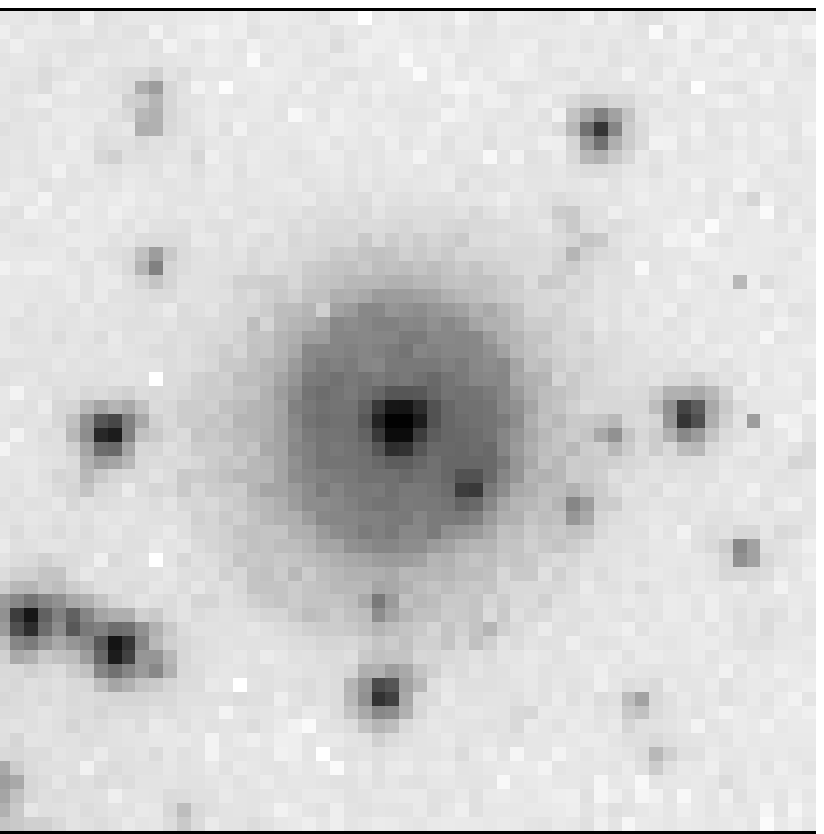}{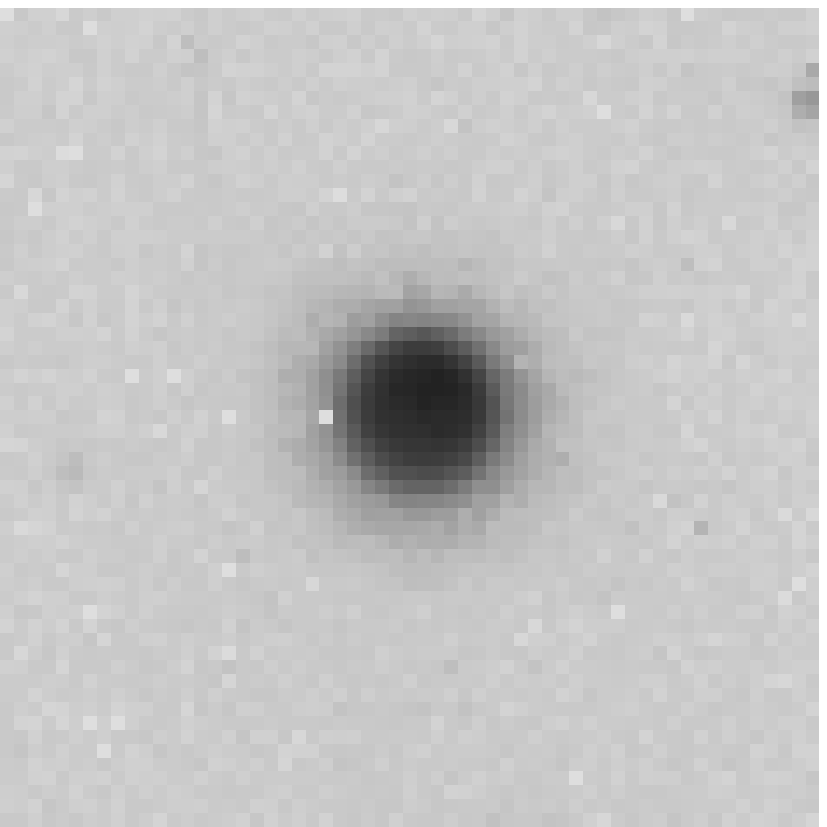}
\caption{Round PNs: LMC J~33 and SMC SP~34.} 
\end{figure}

\subsection{PN morphology}

The morphology of Galactic PNs has been studied rather thoroughly in the past
decade, and it has been found that the morphological types correlate with
the PN progenitor's evolutionary history, and the stellar mass. There is strong evidence
that asymmetric (e. g., bipolar) PNs are the progeny of the massive
AGB progenitors (3-8 M$_{\odot}$). Bipolar PNs are nitrogen enriched and carbon poor
(Stanghellini et al. 2002b). The analysis of the morphological 
types and their distribution in a PN population is then very useful to infer
the age and history of a given stellar sample.

Galactic PNs have been classified as round, elliptical, bipolar (and quadrupolar), bipolar
core (those bipolar PNs whose lobes are too faint to be detected, but whose 
equatorial ring is very evident), and point-symmetric.
The majority of Galactic PNs are elliptical, but the actual number of
bipolars could be underestimated,
given that they typically lie in the Galactic plane (i.e., they may suffer high 
reddening).

In Figures 1 through 3 we show samplers of the most common morphological types
of Magellanic PNs, round, elliptical, and bipolar. These PNs are 
more than 50 times farther away than the typical galactic PNs, yet the major 
morphological features are easily recognized, as are the location of their
CSs, when visible. 

PNs in the Clouds, when spatially resolved, show the same admixture of morphological types 
than the Galactic PNs. While we do not attempt a statistical comparison of the
MC and Galactic PN morphological types, given the selection effects that hamper
Galactic PNs, we can meaningfully compare the LMC and SMC samples. Both
samples suffer from low field extinction, and they have been preselected in
more or less the same way.

\begin{figure}
\epsscale{.40}
\plottwo{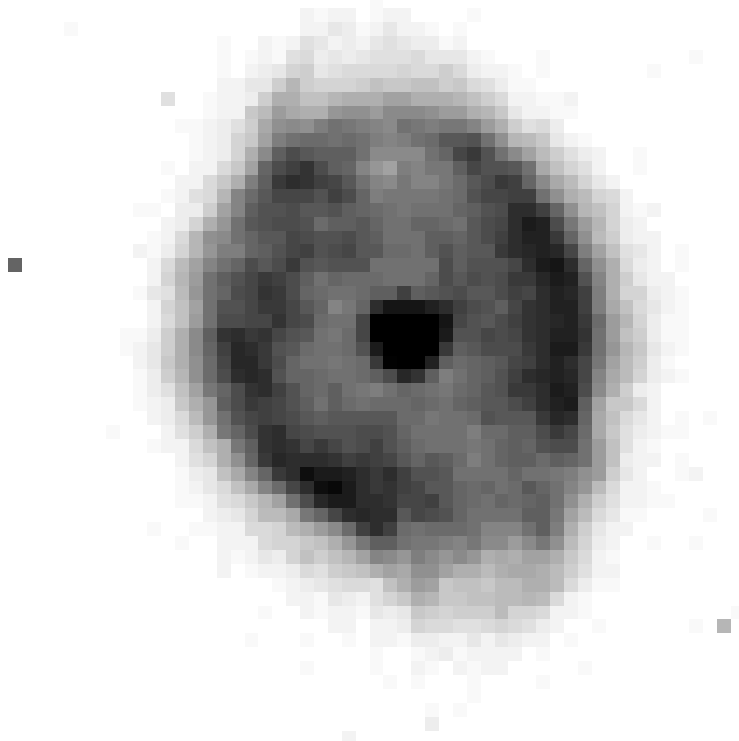}{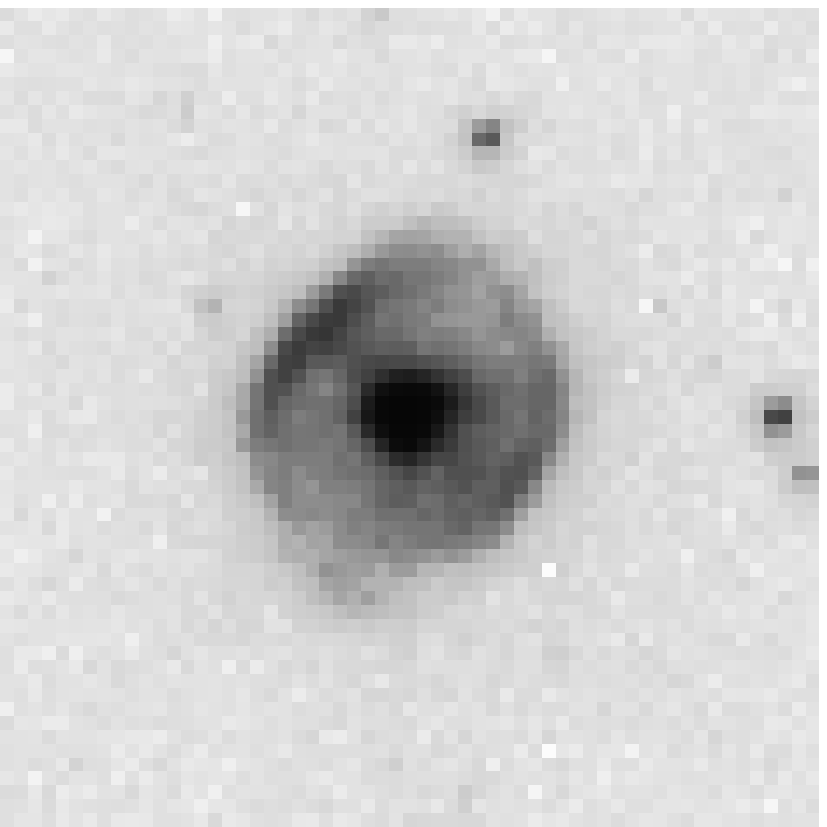}
\caption{Elliptical PNs: LMC SMP~101 and SMC MG~8.}
\end{figure}

The results of the morphological distribution in the Clouds is summarized in Table 1.
Together with the percentage in each morphological class, we give
the total of symmetric (round and elliptical) and asymmetric (bipolar and bipolar
core) PNs. One striking difference between the two distributions is that the fraction of
bipolar PNs in the LMC is almost six times that of the SMC. Bipolar PNs are easily 
recognized, thus this is a sound result. If we add to the asymmetric PN count the 
bipolar core PNs, we obtain that half of the LMC PNs are asymmetric, while only
a third of the SMC PNs are asymmetric. Observational biases play in the same way for the
two samples.

\begin{table}
\label{tab2}  
\caption{Morphological distribution}
\begin{center}
\begin{tabular}{lll}
&&\\
Morphological type& $\%$ LMC&  $\%$ SMC\\
&&\\  
Round (R)&		29&	35\\
Elliptical (E)&		17&	29\\
R+E (symm.)&		46&	64\\
Bipolar (B)&		34&	6\\
Bipolar core (BC)&	17&	24\\
B+BC (asymm.)&		51&	30\\
Point-symmetric (P)&	3&	6\\	

    \end{tabular}

\end{center}
\end{table}

What insight can we get from the morphological results? First of all, it is
clear that the set of processes that are involved in the formation of the
different PN shapes are at work {\it in all galaxies where morphology has been
studied}. Second, the SMC environment may disfavor the onset
of bipolarity in PNs. Otherwise, the different morphological statistics 
may indicate different populations of stellar progenitors in the two Clouds.
While it seems reasonable to conclude that a low metallicity environment 
is unfavorable to bipolar evolution, the exact causes have not been studied
yet. A detailed study of metallicity and mass loss may clarify this 
point. On the other hand, the different morphological statistics may simply 
be related to a lower average stellar mass of the PN progenitors in the SMC.
If this was the case, we should observe also lower CS masses in the
SMC PNs than in the LMC PNs. Our preliminary measurements seem to indicate that
this is also the case (Villaver, Stanghellini, \& Shaw, in preparation).

\subsection{Surface brightness evolution}

The surface brightness of LMC and SMC PNs correlates with the photometric
radius (Stanghellini et al. 2002a, 2003a).
The surface brightness-
photometric radius relation is tight in all spectral lines, with the 
exception of the [N II] emission line, where a larger spread is present,
particularly for bipolar PNs. A possible factor is the
larger range of nitrogen abundances in bipolar and BC PNs. 
The surface brightness-
photometric radius relations hold
only in the cases in which the nebular density $N_{\rm e}$ is smaller
than the  critical density, $N_{\rm crit}$ (the density at which the
collisional de-excitation rate balances the radiative  transition rate).

A good eye-fit to the surface brightness-
photometric radius relation is SB $\propto$ R$_{\rm phot}^{-3}$. This relation
can be reproduced via hydrodynamic modeling of evolving PNs and their
CSs (Villaver \& Stanghellini, in preparation). The surface brightness-
photometric radius relation in the light of H$\alpha$ (or H$\beta$) is tight enough that
it can be used to set the distance scale for Galactic PNs with intrinsic
uncertainties of the order of 30$\%$ or less (Stanghellini et al. in preparation), 
while the current calibration of the Galactic PN distance scales carry errors
of the order of 50$\%$ or more.

In the surface brightness-
photometric radius relation we note that the symmetric (round and elliptical) PNs tend to
cluster at high surface brightness and low radii, while the asymmetric
PNs occupy the lower right parts of the diagrams. This separation can be
interpreted with a slower evolutionary rate for the symmetric PNs, 
which agrees with the idea that symmetric PNs derive from lower mass progenitors
(Stanghellini, Corradi, \& Schwarz 1993, Stanghellini et al. 2002b).

\begin{figure}
\epsscale{.40}
\plottwo{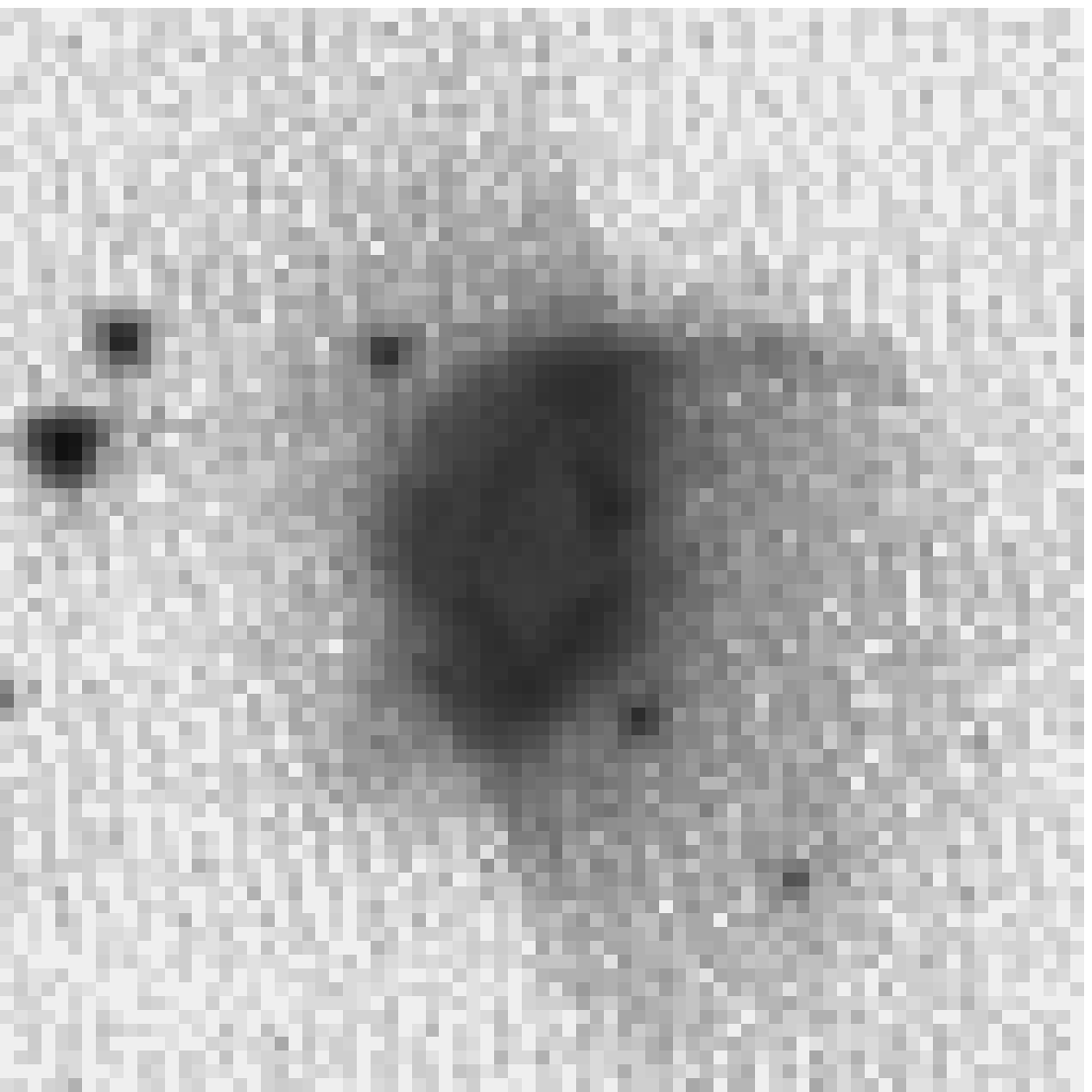}{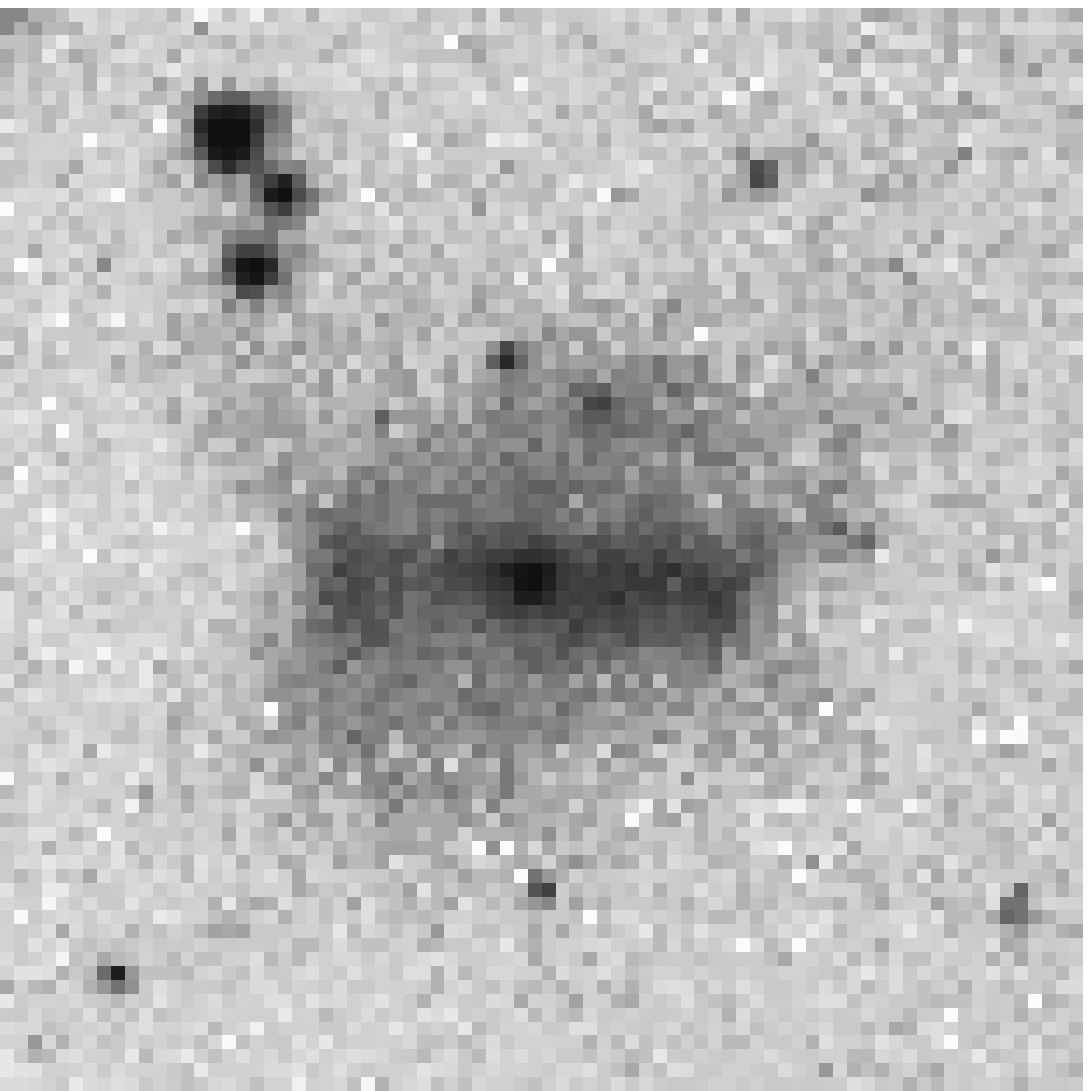}
\caption{Bipolar PNs: LMC SMP~91 and SMC MA~1682.} \label{}
\end{figure}

\subsection{The [O III]/H$\beta$ distribution}

In Figure 4 (left panel) we plot a histogram of the ratio of (reddening-corrected)
fluxes of the [O III] $\lambda$5007 and H$\beta$ lines 
for the PNs of the SMC and the LMC.  The median of the 
SMC distribution is a factor of two lower than for the corresponding LMC 
distribution (
$<$[O III]/H$\beta$$>_{\rm SMC}$ = 5.7 $\pm$ 2.5 and $<$[O III]/H$\beta$$>_{\rm LMC}$ = 9.4 $\pm$ 3.1).  
This result is free of object selection 
biases since both sets of targets were chosen in much the same way.

The [O III]/H$\beta$ emissivity ratio is physically scaled linearly with 
the O/H abundance and the fractional ionization of O$\rm^{++}$. Also it 
depends exponentially on the local electron excitation temperature, 
T$_e$(O$\rm^{++}$) since electron collisions on the high-energy tail of 
the free energy distribution excite the transition.  Of course, 
T$_e$(O$\rm^{++})$ depends on O/H and O$\rm^{++}$/O as well.  So 
interpreting the differences between the [O III]/H$\beta$ ratios of the 
SMC and the LMC is best done using ionization models. 

Our Cloudy (Ferland 1996) models explore the major line emission
in a set of Galactic, LMC, and SMC models with same gas density (1000 cm$^{-3}$)
and different metallicities, adequately chosen to represent 
the {\it average} nebula in each studied galaxy, as explained in Stanghellini et
al. (2003a).
The stellar ionizing spectrum 
is assumed to be a blackbody with temperatures and luminosities from the 
H-burning evolutionary tracks for the appropriate galaxian population
by Vassiliadis and Wood (1994).

In Figure 4 (right panel) we show the line intensity relative to H$\beta$
for the major coolants in the SMC, LMC, and Galactic PNs, versus the oxygen abundance,
as derived from our simplified Cloudy models. 
While we have calculated the evolution of these intensity ratios following the
evolution of the CS from
the early post-AGB phase to the white dwarf stage, 
we only plot here the flux ratios corresponding to the models
with the highest temperature, for each PN
composition.  In general, our target selection tends to favor targets with hottest 
CSs: T$_{eff} \ge 50,000$ K both in the SMC and in the LMC, thus
the set of high-temperature models is the most adequate to reproduce the 
observations for LMC and SMC PNs.

The cooling processes that determine T$_e$(O$\rm^{++})$ in the SMC, LMC 
and Galactic PNs are noteworthy.  In the Galaxy the primary coolants of 
PNs with hot CSs are the optical forbidden lines of [O III]
$\lambda$5007 and other lines of O$\rm^+$ and O$\rm^{++}$.  However, 
in environments in which O/H is as low as in the SMC, the primary 
coolants may become ultraviolet intercombination lines of 
C$\rm^+$ and C$\rm^{++}$.  
The simple models described here seem to reproduce very well the optical
flux ratios of PNs in the Magellanic Clouds.
It will be interesting to confirm these 
predictions with future UV observations.

\begin{figure}
\epsscale{.40}
\plottwo{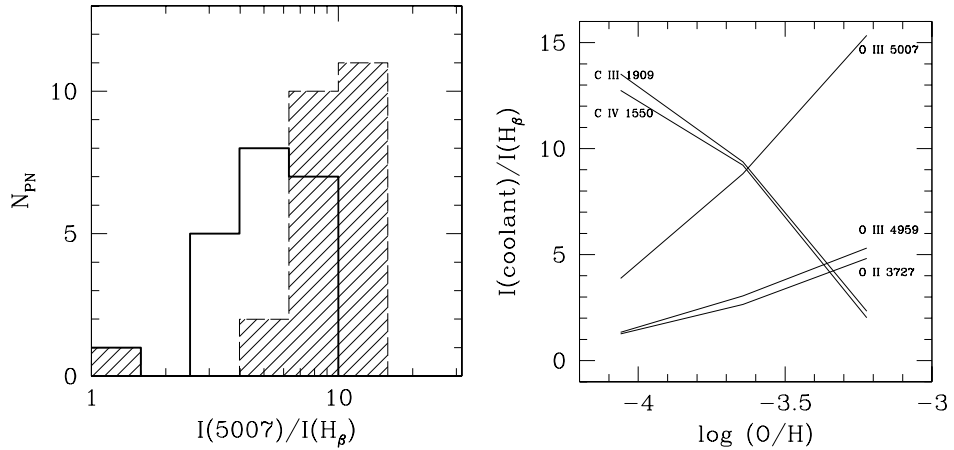}{fig4.epsi}
\caption{Left: distribution of the [O III] $\lambda$5007 over H$\beta$ intensity ratios
in the SMC (thick histogram) and LMC (shaded histogram) PNs;
right: intensity ratios of the major PN coolants over H$\beta$, versus oxygen
abundances (from Stanghellini et al. 2003a).}
\end{figure}
 
\section{Summary}

Magellanic PNs are ideal probes to study stellar evolution and populations of 
low- and intermediate-mass stars. The use of the 
{\it HST} is fundamental for determining the PN shapes, the radii, and
also to detect the CSs. Furthermore, only with the use of spatially
resolved images one can identify the LMC and SMC PNs unambiguously,
without the accidental inclusion of compact H II regions in the PN samples. 

We have presented some of the results derived from our {\it HST}
programs.
We found that PNs have the same morphological types in the Galaxy, the LMC, and
the SMC. We also found that the distribution of the morphological types
is noticeably different in the SMC and the LMC, and that the LMC seems to be
populated by PNs whose progenitors are, on average, more massive.

An empirical relation between the nebular radii and the surface brightness
is found to hold in both SMC and LMC PNs, independent of morphological type.
The relation, once calibrated, will be used to determine the distance scale for
Galactic extended PNs. 

The PN cooling is affected by metallicity, and it seems that the [O III] $\lambda$5007
emission is not always the ideal line to detect bright PNs in all Galaxies,
since the strongest cooling lines in very low metallicity PNs
seem to be the UV C III] (and C IV]) semiforbidden emission. While the
[O III] luminosity functions for the LMC and the SMC PNs are available from the 
ground, only {\it HST} can unambiguosly determine whether the selected
objects are indeed PNs, or are instead H II regions related to young
stellar clusters. Since the ambiguity is metallicity-dependent (see Stanghellini
et al. 2003b), the result found here is extremely novel.

The observed Magellanic CSs that we did not discuss here in detail
constitute the first sizable sample of CS beyond the Milky Way that
has been directly observed. While we found only marginal differences
between the LMC and the SMC median CSs masses of the CSs, we need to enlarge the sample of CS 
whose masses can be reliably measured, given the importance of knowing 
initial-to final-mass relation in different metallicity environments
(Villaver et al. 2003; Villaver et al. in preparation).

\acknowledgements
My {\it Magellanic Cloud Planetary Nebulae}
collaborators, Dick Shaw, Eva Villaver, Chris Blades, and Bruce Balick,
are warmly acknowledged for their contributions to the 
project.

%
%
%
%


\end{document}